\documentclass[aps,prl,twocolumn,floatfix]{revtex4}

\usepackage[mathscr]{eucal}
\usepackage{amsfonts}
\usepackage{hyperref}
\usepackage{amsmath,amssymb}
\usepackage{color}
\usepackage{graphics}
\usepackage {epsfig}
\usepackage{psfrag}
\usepackage{graphicx}

\begin{document}

\title{Cluster state generation with atomic ensembles
via the dipole blockade mechanism}
\author{Marcin Zwierz}\email{php07mz@sheffield.ac.uk}
\author{Pieter Kok}
\affiliation{Department of Physics and Astronomy, University of
Sheffield, Hounsfield Road, Sheffield, S3 7RH, UK}

\date{\today}

\begin{abstract}
We present a new scheme for cluster states generation based on atomic
ensembles and the dipole blockade mechanism. The protocol requires identical
single photon sources, one ensemble per physical qubit, and regular
photodetectors. The general entangling procedure is presented, as well as a
procedure that generates $Q$-qubit GHZ states with probability
$p\sim\eta^{Q/2}$, where $\eta$ is the combined detection and source
efficiency. This is significantly more efficient than any known robust
probabilistic entangling operation. The GHZ states form
the basic building block for universal cluster states ---
a resource for the one-way quantum computer.
\end{abstract}

\pacs{}
\maketitle

\noindent
\paragraph{Introduction.}
The construction of a quantum computer is an important goal of
modern physics, and one possible implementation is via atomic ensembles:
The quantum state of the ensemble can be coherently manipulated with
light, and the decoherence of the quantum information can be highly
suppressed \cite{fleisch,lukin,duan,barrett,hammerer}. It is therefore possible
to define a ``good'' qubit in an atomic ensemble, and the question
remains how to implement the entangling operations between the qubits
that enable universal quantum computation. It suffices to create a large
entangled multi-qubit resource ---the cluster state--- after which the
entire computation proceeds via single-qubit measurements
\cite{rauss,hein}. Here, we show how to create these cluster states
using the dipole blockade mechanism. The protocol requires identical
single photon sources, one ensemble per physical qubit, and regular
photodetectors. We present a general entangling procedure, as well as a
procedure that generates $Q$-qubit GHZ states with probability
$p\sim\eta^{Q/2}$, where $\eta$ is the combined detection and source
efficiency. This is significantly more efficient than any known robust
probabilistic entangling operation \cite{lim,kok}. The GHZ states form
the basic building block for universal cluster states.

The physical mechanism that is central to our proposal is the dipole
blockade mechanism in the atomic ensemble: An optical pulse resonant
with a transition to a Rydberg state will create a Rydberg atom with a
very large dipole moment. When the atoms in the ensemble are
sufficiently close, the dipole interaction between the Rydberg atom and
the other atoms will cause a shift in the Rydberg transition energy of
the other atoms. Therefore, the optical pulse becomes off-resonant with
the other atoms, and the ensemble is transparent to the pulse. This
mechanism prevents populating states of atomic ensembles with two or
more atoms excited to the Rydberg level \cite{lukin,johnson}.

The range and quality of the dipole interaction has been studied
extensively: Walker and Saffman analyzed the primary errors that enter
the blockade process \cite{walkerbloc,van}. For Rubidium atoms with
principal quantum number $n=70$, the blockade energy shift is
approximately 1~MHz. Hence, a strong and reliable blockade is possible
for two atoms with separation up to $\sim10~\mu$m \cite{walkerbloc}.
Moreover, decoherence associated with spontaneous emission from
long-lived Rydberg states can be quite low ($\sim$ 1 ms). The dipole
blockade mechanism can be used to build fast quantum gates, i.e., a two
qubit phase gate \cite{jaksch,lukingate,brion}. The long-range
dipole-dipole interaction between atoms can be employed to realize a
universal phase gate between pairs of a single-photon pulses
\cite{friedler,mohapatra,petrosyan}.

\begin{figure}
\includegraphics[scale=0.60]{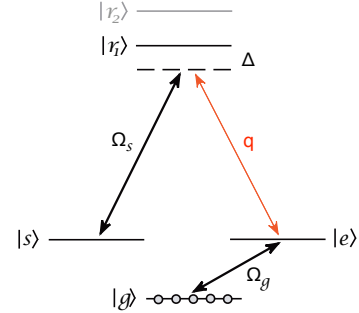}
\caption{Relevant atomic level structure with allowed atomic
transitions. States $|g \rangle$, $|e \rangle$ and $|s \rangle$ can be
realized by the electronic low-lying states of alkali atoms. The
transition between states $|g \rangle$ and $|s \rangle$ is always
dipole-forbidden. The state $|g \rangle$ is coupled to the state $|e \rangle$
through a classical field $\Omega_{g}$. A second
classical field $\Omega_{s}$ is at resonance with the transition between the highly
excited Rydberg level $| r_{1} \rangle$ and the state $| s \rangle$. States $|e
\rangle$ and $|r_{1} \rangle$ are coupled via a quantum field. $|r_{2}
\rangle$ is an auxiliary Rydberg level used in single-qubit operations. \label{levels}}
\end{figure}

\begin{figure}
\includegraphics[scale=0.45]{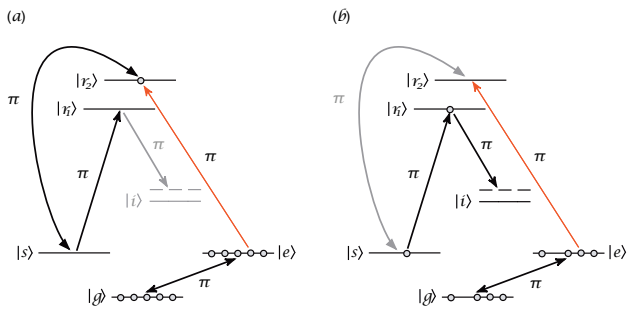}
\caption{The bit flip operation ($X$) and the Hadamard gate ($H$). First, two
$\pi$-pulses are applied to the transitions $| g \rangle \leftrightarrow | e \rangle$ and $|
s \rangle \leftrightarrow | r_{1} \rangle$. Then we send a
pulse resonant with the transition $| e \rangle \leftrightarrow |
r_{2} \rangle$. Finally, two pulses couple levels $| r_{1} \rangle$
and  $| e \rangle$ (STIRAP through the level $| i \rangle$) and levels
$| r_{2} \rangle$ and $| s \rangle$. We apply the same sequence of
classical pulses in both cases. (a) $| 0 \rangle_{L} \rightarrow | 1
\rangle_{L}$ After a first two pulses no atom is present in the state $|
r_{1} \rangle$. Therefore, one atom may be excited to the state $| r_{2}
\rangle$ and then transferred to the state $| s \rangle \equiv | 1
\rangle_{L}$. (b) $| 1 \rangle_{L} \rightarrow | 0 \rangle_{L}$ Now the
presence of one atom in the state $| r_{1} \rangle$ blocks the excitation
of another atom to the state $| r_{2} \rangle$. The STIRAP procedure
transfers a single atom from state $| r_{1} \rangle$ to state $| e
\rangle$. In this setup, the Hadamard gate can be realized by a $\pi/2$-pulse
between Rydberg states $| r_{1} \rangle$ and $| r_{2} \rangle$. A
single excitation can be transferred to the storage states $| g \rangle$
and $| s \rangle$. More details about single-qubit operations can be
found in Ref. \cite{zwierz}. \label{X&H}}
\end{figure}

\paragraph{Protocol.}
Each qubit is represented by a spatially separated atomic ensemble. The
atoms in each ensemble have three lower, long lived energy states $|g
\rangle$, $|e \rangle$ and $|s \rangle$ (see Fig.~\ref{levels}). The qubit
states in an ensemble of $N$ atoms are:
\begin{eqnarray}
|0\rangle_{L} &\equiv& |g\rangle = |g_{1},g_{2}, \ldots ,g_{N}\rangle\, , \\
|1\rangle_{L} &\equiv& |s\rangle = \frac{1}{\sqrt{N}} \sum_{j=1}^{N}
|g_{1},g_{2}, \ldots ,s_{j}, \ldots ,g_{N}\rangle\, .
\end{eqnarray}
The states $|e \rangle$ and $|r_{1} \rangle$ participate in the
interaction part of the scheme. Levels $|g \rangle$ and $|s \rangle$
play the role of the storage states. Single-qubit operations are
realized by means of classical optical pulses and the dipole blockade
mechanism. An arbitrary phase gate $\Phi(\phi) = \exp(-i\phi Z/2)$ is
realized by a detuned optical pulse applied to the transition between
$|s\rangle$ and an auxiliary level $|a\rangle$ (not shown). The Pauli
$X$ operation (the bit flip) and the Hadamard gate $H$ are shown in
Fig.~\ref{X&H}. The gates $\Phi(\phi)$, $X$, and $H$ generate all
single-qubit operations.  The readout of a qubit is based on resonance
fluorescence. If the measurement gives no fluorescence photons, the
qubit is in  $|0\rangle_{L}$. Otherwise, the state of the qubit is
projected onto $|1\rangle_{L}$.

\begin{figure}
\includegraphics[scale=0.60]{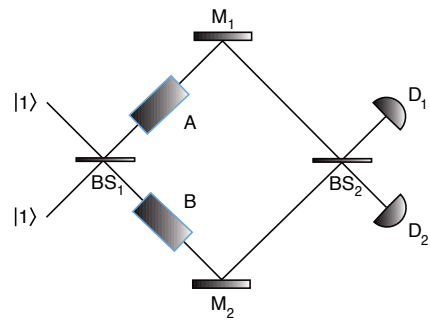}
\caption{Diagram of the protocol. We send an entangled pair of photons in
the state $|\phi_{light}\rangle=\frac{i}{\sqrt{2}} (|02\rangle +
|20\rangle)$ into ensembles A and B. The photons interact with atomic
vapours: one and only one alkali atom in the ensemble is excited by one
of the photons to a Rydberg state $|r_{1}\rangle$. Absorption of the
second photon is prohibited by the dipole blockade mechanism. After $BS_{2}$,
a state of ensembles-light system has the following form $|\phi_{out}\rangle=\frac{i}{\sqrt{2}}
(|\psi^{+}\rangle |01\rangle + |\psi^{-}\rangle |10\rangle)$, where
$|\psi^{\pm}\rangle = \frac{1}{\sqrt{2}}(|r_{1}e\rangle \pm i
|er_{1}\rangle)$. Detection of a single photon will leave the atomic ensembles
entangled. \label{scheme}}
\end{figure}

The entangling operation is constructed as follows: Two atomic ensembles
are placed in the arms of a Mach-Zehnder interferometer (see
Fig.~\ref{scheme}). Initially, we prepare each ensemble $A$ and $B$ in
the state $|\phi_{A,B}\rangle = |e\rangle \equiv |e_{1},e_{2}, \ldots
,e_{N}\rangle$. Two indistinguishable photons enter each input mode of
the interferometer, and due to the Hong-Ou-Mandel (HOM) effect, after
the first beam splitter ($BS_{1}$) the two photons are in the state
$|\phi_{light}\rangle=|11\rangle \xrightarrow{BS_{1}} \frac{i}{\sqrt{2}}
(|02\rangle + |20\rangle)$ where $|0\rangle$ and $|2\rangle$ denote the
vacuum and a two-photon state, respectively. After $BS_{1}$ the photons
interact with the atomic ensembles: one and only one atom in the
ensemble is excited by one of the photons to a Rydberg state
$|r_{1}\rangle$, and the absorption of the second photon is prohibited
by the dipole blockade mechanism. The total state of a ensemble-light
system after interaction is given by:
\begin{equation}
|\phi_{int}\rangle=\frac{i}{\sqrt{2}} (|er_{1}\rangle |01\rangle +
|r_{1}e\rangle |10\rangle).
\end{equation}
After the second beam splitter ($BS_{2}$), the total state is
\begin{equation}
|\phi_{out}\rangle=\frac{i}{\sqrt{2}} (|\psi^{+}\rangle |01\rangle +
|\psi^{-}\rangle |10\rangle),
\end{equation}
where $|\psi^{\pm}\rangle = \frac{1}{\sqrt{2}}(|r_{1}e\rangle \pm i
|er_{1}\rangle)$. Conditional on a single photodetector click, the
ensembles are projected onto a maximally entangled state. After
establishing entanglement, the qubits are transferred to their
computational basis states $|0\rangle_{L} \equiv |g\rangle$ and
$|1\rangle_{L} \equiv |s\rangle$ by classical optical pulses
$\Omega_{g}$ and $\Omega_{s}$. This means that ideally every run of the
procedure would give an entangled state of ensembles with success
probability $p = \eta$, where $\eta$ is the combined detection and source
efficiency. This is a significant improvement compared to the
success probability $p = \eta^2/2$ of the double heralding protocol in
Ref.~\cite{kok}.

\begin{figure}
\includegraphics[scale=0.60]{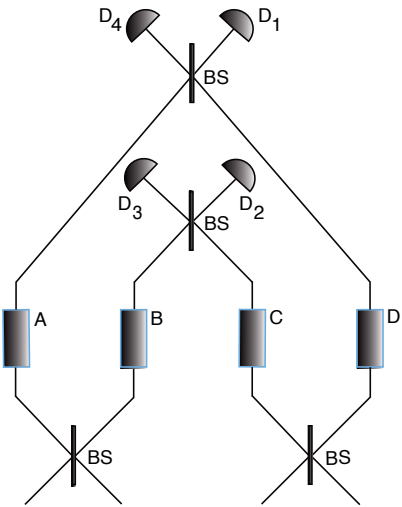}
\caption{The scheme for creating a 4-qubit GHZ state. The four ensembles
$A$, $B$, $C$ and $D$ are prepared in the state
$|\phi_{ABCD}\rangle=|eeee\rangle$. Four indistinguishable photons are sent into
the beam splitters. The interaction of photons with the atomic
vapours is followed by the beamsplitters and four photodetectors.
Conditional on photodetector clicks at the photodetector $(D_{1},D_{2})$,
$(D_{1},D_{3})$, $(D_{4},D_{2})$ or $(D_{4}, D_{3})$,
a state of the four qubits is projected onto the 4-qubit GHZ
state (up to phase correcting operations) with success probability $p =
\eta^2/2$. \label{GHZ}}
\end{figure}

\paragraph{GHZ and cluster states.}
The entangling operation can be used to efficiently create arbitrary
cluster states, including universal resource states for quantum
computing. However, a modification of the entangling procedure yields an
even more dramatic improvement in the efficiency of cluster state
generation. By arranging the ensembles in a four-mode interferometer as
shown in Fig.~\ref{GHZ}, the detection of two photons will create a
four-qubit GHZ state in a single shot. Moreover, since only two photons
are detected, the protocol is relatively insensitive to detector losses.
The success probability is $p = \eta^2/2$. Higher GHZ states can
be created by a straightforward extension. A subsequent cluster states
are generated with probability $p = \eta^{Q/2} (Q-2)/2^{Q-2}$ where $Q =
4, 6, \ldots$ is the number of the qubits.

The efficiently generated large GHZ states may serve as a building block
for creating arbitrary cluster states. By entangling small clusters with
the above entangling procedure, large cluster states can be constructed.
A single photon applied to a pair of qubits (each from two different
4-qubit cluster states) followed by a single photodetector click creates
a 8-qubit cluster state with success probability $p = \eta'/8$. This
procedure can be repeated in an efficient manner \cite{kieling}. In case
of failure, the two qubits that participated in linking are measured in
the computational basis, and the rest of the cluster state is recycled.

\paragraph{Errors, decoherence mechanisms and fidelity.}
The dominant errors and decoherence mechanisms are the coincident event
in the HOM effect, the spontaneous emission rate of the Rydberg state,
the black-body transfer rate (to other Rydberg states), the atomic
collision rate, the doubly-excited Rydberg states and singly-excited
states outside the desired two-level system, no absorption event, and
the inefficiency and the dark count rate of the photodetectors. These
errors are discussed in the next section of article. Considering the time scale
of the protocol, the entangling procedure is mostly affected by the no
absorption event and inefficiency of the photodetectors (we assume that
the coincident event rate in the HOM effect is negligible). In presence of above noise and
decoherence mechanisms, the final state of the system conditional on a
single photodetector click is given by
\begin{equation}
  \rho_{fin} = (1 - 2\varepsilon) \; |\psi^{\pm}
\rangle \langle \psi^{\pm}| + 2\varepsilon \;
\rho_{noise} + \mathcal{O}(\varepsilon^2)\, ,
\end{equation}
where $|\psi^{\pm}\rangle = \frac{1}{\sqrt{2}}(|sg\rangle \pm i
|gs\rangle)$ and $\varepsilon = 1-P_{abs}$.
$\rho_{noise}$ denotes the unwanted terms in the state of the two
ensembles. The source efficiency does not affect the fidelity of the final state,
just lowers the success probability.
After taking into account all dominant error mechanism, the fidelity of
the prepared entangled state is $F = \langle \psi^{\pm} | \rho_{fin} |
\psi^{\pm} \rangle \cong 0.982$, which is close to current
fault-tolerant thresholds \cite{nielsen}.

\paragraph{Experimental implementation.}
Let us analyze in more detail mentioned dominant error and decoherence mechanisms.
First, consider the coincident events in the HOM effect. The single
indistinguishable photons that recombine at the first beam splitter
($BS_{1}$) can be generated by means of a spontaneous parametric
down-conversion (SPDC) process. In general, successful generation of the
entangled state of light depends on the proper setup, where both photons
recombine at $BS_{1}$ at the same time. In a
recent experiment, the coincident event in the HOM effect happens with a
rate of 1500 counts/s \cite{kim}. We assume that the rate of coincident
events is negligible comparing to the time scale of the protocol which
is $t \cong 5\mu$s.

Assume that an atomic vapour consists of 300 $^{87}$Rb atoms placed in
the far-off-resonant optical trap or magneto-optical trap (MOT). The
atomic levels $|g \rangle$, $|e \rangle$ and $|r \rangle$ may correspond
to $5S_{1/2}$, $5P_{3/2}$ and $43D_{5/2}$ or $58D_{3/2}$, respectively.
State  $|s \rangle$ corresponds to the long lived, lower energy level.
The spatial distribution of an atomic cloud is a quasi one-dimensional
ensemble with probability density $P(z) = (2 \pi \sigma^2)^{-1/2}
\mbox{exp}(-z^2/2 \sigma^2)$ where $\sigma = 3.0$ $\mu\mbox{m}$. Atomic
vapours described with quasi  one-dimensional probability density have
been demonstrated experimentally \cite{johnson}.\\
When a protocol is based on a quantum optical system, its performance is
limited by the inefficiency and the dark count rate of the
photodetectors. The dark count rate of the modern photodetector can be
as low as 20 Hz and efficiency reaches $\eta \approx 30\%$ for
wavelengths around 480 nm. The probability of the dark count is
$P_{dc} = 1 - \mbox{exp}(-\gamma_{dc} t/p_{success} ) \cong 5 \ 10^{-4}$.
In general, the probability of the dark count is negligible for $p_{success} > \gamma_{dc} t$.\\
Since the length of the atomic ensemble needs to be of order of $\mu$m,
the most important source of errors is the lack of absorption event. The
probability of an absorption of a single photon by an atomic ensemble is
given by $P_{abs} \cong  1 - e^{-N_{i} \sigma_{0} /A}$ with $N_{i}$ the
number of atoms in the interaction region, $\sigma_{0} = 3 \lambda^2/(2
\pi)$ and $A = \pi w^2_{0}$ \cite{walkerabs}. With  $\lambda_{43D} =
485.766$ nm, $\lambda_{58D} = 485.081$ nm and $w_{0} \approx \pi
\lambda$, the probability of an absorption for both Rydberg states is
$P_{abs} \cong 0.989$.\\
The probability of doubly-excited Rydberg states (absorption of both
photons by an ensemble) depends on the quality of the dipole blockade
and is given by the following expression $P_{2} =(N-1) g^2_{N}/2 N
B^2$, where $g_{N} = \sqrt{N} g_{0}$ and $B$ is the mean
blockade shift \cite{walkerbloc}. For $43D_{5/2}$ and $58D_{3/2}$ mean
blockade shift $B = 0.25 \ \mbox{MHz}$ and $B = 2.9 \ \mbox{MHz}$ in a
trap with $\sigma = 3.0$  $\mu\mbox{m}$, respectively \cite{walkerbloc}.
Hence, the probability of doubly-excited states for $43D_{5/2}$ level
is $P_{2} \cong 0.26 $ and for $58D_{3/2}$ level
is $P_{2} \cong 0.57 \ 10^{-3}$. The probability of doubly-excited states and
singly-excited states outside the desired two-level system are similar.
Above probabilities are given for the worst case scenario when the separation of
atoms is maximal and the dipole-dipole interaction is of the weakest, van der Waals type.\\
The spontaneous emission from the Rydberg state and the black-body
transfer (to other Rydberg states) occur with rate of order $10^3$ Hz,
and is negligible, since after successful entanglement preparation the state
of ensemble is stored in the long lived atomic states  $|g \rangle$ and
$|s \rangle$. Exact values of these rates are given in Ref.
\cite{kim,day}. The atomic collision rate is given by
$\tau^{-1}_{col}\approx n \sigma_{col}/\sqrt{M/3 k_{B} T}$ with $n$ the
number density of atoms, $\sigma_{col}$ the collisional cross section
($\sim 10^{-14}$ $\mbox{cm}^2$), $M$ the atomic mass, $k_{B}$ the
Boltzmann's constant, and $T$ the temperature  \cite{james}. Assuming a
vapour with the number density of atoms of order $10^{12}$  cm$^{-3}$
and with the temperature of about $10^{-3}$ K (which
implies negligible Doppler broadening), the atomic collision rate is as
low as 2 Hz. Moreover, with a sufficiently large energy difference
between states $|g\rangle$ and $|s\rangle$ a single collision is not
likely to affect the qubit.

\paragraph{Conclusions.}
In conclusion, we presented a new scheme for cluster state generation
based on atomic ensembles and the dipole blockade mechanism. The
protocol consists of single-photon sources, atomic ensembles, and
realistic photodetectors. The protocol generates GHZ states with
probability $p\sim \eta^{Q/2}$, where $Q$ is the number of the qubits,
and high fidelity $F \cong 0.982$. The protocol is more efficient than
any previously proposed probabilistic scheme with realistic
photodetectors and sources. In general, number-resolution photodetectors
are not required. \\

\begin{acknowledgments}
We thank Charles Adams and Matthew Jones for helpful discussion. This
work was supported by the White Rose Foundation.\\
\end{acknowledgments}


\begin{thebibliography}{99}

\bibitem{fleisch} Fleischhauer, M., Imamo\={g}lu, A. \& Marangos, J. P.
Electromagnetically induced transparency: Optics in coherent media.
\textit{Rev. Mod. Phys.} \textbf{77,} 633 (2005).

\bibitem{lukin} Lukin, M. D. Colloquium: Trapping and manipulating
photon states in atomic ensembles. \textit{Rev. Mod. Phys.} \textbf{75,}
457 (2003).

\bibitem{duan} Duan, L.-M., Lukin, M. D., Cirac, J. I. \& Zoller, P.
Long-distance quantum communication with atomic ensembles and linear
optics. \textit{Nature} \textbf{414,} 413 (2001).

\bibitem{barrett} Barrett, S. D., Rohde, P. P. \& Stace, T. M.
Scalable quantum computing with atomic ensembles. Preprint at
http://arxiv.org/abs/0804.0962v1 (2008).

\bibitem{hammerer} Hammerer, K., Sorensen, A. S. \& Polzik, E. S.
Quantum interface between light and atomic ensembles. Preprint at
http://arxiv.org/abs/0807.3358v1 (2008).

\bibitem{rauss} Raussendorf, R., Browne, D. E. \& Briegel, H. J.
Measurement-based quantum computation on cluster states. \textit{Phys.
Rev. A} \textbf{68,} 022312 (2003).

\bibitem{hein} Hein, M., Eisert, J. \& Briegel, H. J. Multiparty
entanglement in graph states. \textit{Phys. Rev. A} \textbf{69,} 062311
(2004).

\bibitem{kok} Barrett, S. D. \& Kok, P.  Efficient high-fidelity quantum
computation using matter qubits and linear optics. \textit{Phys. Rev. A}
\textbf{71,} 060310(R) (2005).

\bibitem{lim} Lim, Y. L., Barrett, S. D., Beige, A., Kok, P. \& Kwek,
L. C.  Repeat-until-success quantum computing using stationary and flying qubits.
\textit{Phys. Rev. A} \textbf{73}, 012304 (2006).

\bibitem{johnson} Johnson, T. A. \textit{et al.} Rabi flopping between
ground and Rydberg states with dipole-dipole atomic interactions.
Preprint at http://arxiv.org/abs/0711.0401v1 (2007).

\bibitem{walkerbloc} Walker, T. G. \& Saffman, M.  Consequences of
Zeeman degeneracy for the van der Waals blockade between Rydberg atoms.
\textit{Phys. Rev. A} \textbf{77,} 032723 (2008).

\bibitem{van} van Ditzhuijzen, C. S. E. \textit{et al.} Spatially
resolved observation of dipole-dipole interaction between Rydberg atoms.
Preprint at http://xxx.tau.ac.il/abs/0706.0110v3 (2008).

\bibitem{jaksch} Jaksch, D. \textit{et al.} Fast quantum gates for
neutral atoms. \textit{Phys. Rev. Lett.} \textbf{85,} 2208 (2000).

\bibitem{lukingate} Lukin, M. D. \textit{et al.} Dipole blockade and
quantum information processing in mesoscopic atomic ensembles.
\textit{Phys. Rev. Lett.} \textbf{87,} 037901 (2001).

\bibitem{brion} Brion, E., M{\o}lmer, K. \& Saffman, M.  Quantum
computing with collective ensembles of multilevel systems. \textit{Phys.
Rev. Lett.} \textbf{99,} 260501 (2007).

\bibitem{friedler} Friedler, I., Petrosyan, D., Fleischhauer, M. \&
Kurizki, G. Long-range interactions and entanglement of slow
single-photon pulses. \textit{Phys. Rev. A} \textbf{72,} 043803 (2005).

\bibitem{mohapatra}  Mohapatra, A. K., Jackson, T. R. \& Adams, C. S. Coherent optical detection of highly excited Rydberg states using electromagnetically induced transparency. \textit{Phys. Rev. Lett.} \textbf{98,} 113003 (2007).

\bibitem{petrosyan} Petrosyan, D. \& Fleischhauer, M. Quantum
information processing with single photons and atomic ensembles in
microwave coplanar waveguide resonators. \textit{Phys. Rev. Lett.}
\textbf{100,} 170501 (2008).

\bibitem{zwierz} Zwierz, M. \& Kok, P. In preparation (2008).

\bibitem{kieling} Kieling, K., Gross, D. \& Eisert, J. Cluster state
preparation using gates operating at arbitrary success probabilities.
\textit{New J. Phys.} \textbf{9,} 200 (2007).

\bibitem{nielsen} Nielsen, M. A. \& Chuang, I. L. \textit{Quantum
Computation and Quantum Information} (Cambridge University Press,
Cambridge, 2000).

\bibitem{kim} Kim, H., Kwon, O., Kim, W. \& Kim, T.  Spatial two-photon
interference in a Hong--Ou--Mandel interferometer. \textit{Phys. Rev. A}
\textbf{73,} 023820 (2006).

\bibitem{walkerabs} Saffman, M. \& Walker, T. G. Entangling single- and
N-atom qubits for fast quantum state detection and transmission.
\textit{Phys. Rev. A} \textbf{72,} 042302 (2005).

\bibitem{day} Day, J. O., Brekke, E. \& Walker, T. G. Dynamics of
low-density ultracold Rydberg gases. \textit{Phys. Rev. A} \textbf{77,}
052712 (2008).

\bibitem{james} James, D. F. V. \& Kwiat, P. G. Atomic-vapor-based high
efficiency optical detectors with photon number resolution.
\textit{Phys. Rev. Lett.} \textbf{89,} 183601 (2002).

\end{thebibliography}
\end{document}